\documentclass[%
 reprint,
 amsmath,amssymb,
 aps,
]{revtex4-2}
\usepackage{mathptmx}
\usepackage{etoolbox}
\usepackage{xcolor}
\usepackage{float}
\usepackage{hyperref}
\usepackage{graphicx}
\usepackage{dcolumn}
\usepackage{bm}

\begin{document}


\title{Theoretical prediction of structural stability and superconductivity in T-hexagonal molybdenum dihydrides Monolayer}

\author{Jakkapat Seeyangnok$^{1}$}
 \email{jakkapatjtp@gmail.com} 
\author{Udomsilp Pinsook$^{1}$}%
 \email{Udomsilp.P@Chula.ac.th}
\affiliation{$^{1}$Department of Physics, Faculty of Science,
Chulalongkorn University, Bangkok, Thailand}


\date{\today}

\begin{abstract}
The realization of ambient-pressure, high-temperature superconductivity in hydrogen-rich materials remains a major pursuit in condensed-matter physics. While bulk hydrides require extreme pressures to stabilize, two-dimensional (2D) transition-metal hydrides offer a promising alternative to bypass these compression constraints. In this work, we investigate the structural stability, electronic properties, and phonon-mediated superconductivity of a hexagonal molybdenum dihydride ($\text{MoH}_2$) monolayer using first-principles calculations. Total-energy evaluations reveal that the octahedral $T$-phase is energetically more favorable than the previously reported trigonal prismatic $H$-phase by $0.198\text{ eV}$, establishing the $T$-phase as the true ground-state configuration. Consequently, we systematically evaluate the lattice dynamics and superconducting properties of this ground-state $T\text{-MoH}_2$ monolayer within the frameworks of density functional perturbation theory (DFPT) and the anisotropic Migdal-Eliashberg formalism. The transition metal-hydrogen vibrational networks induce strong electron-phonon coupling (EPC), yielding an integrated coupling parameter of $\lambda = 1.04$. Solving the anisotropic Eliashberg equations predicts a conventional superconducting transition temperature ($T_c$) of $14.4\text{ K}$ at ambient pressure, characterized by a moderately gap distribution ($\Delta = 2.07\text{--}3.01\text{ meV}$ at $5.0\text{ K}$). Our findings highlight the $T\text{-MoH}_2$ monolayer as a structurally, mechanically, and thermally stable platform for exploring low-dimensional conventional superconductivity under ambient conditions.
\end{abstract}

\maketitle

\section{Introduction}
The quest for high-temperature superconductivity within the conventional Bardeen-Cooper-Schrieffer (BCS) framework remains a central frontier in condensed-matter physics. Following Ashcroft's seminal proposal that metallic hydrogen could host room-temperature superconductivity due to its exceptionally high vibrational frequencies and strong electron-phonon coupling (EPC)~\cite{ashcroft2004hydrogen, mcmahon2011high}, significant attention has been directed toward hydrogen-rich materials. This strategy yielded spectacular experimental breakthroughs in high-pressure hydrides, such as $\text{H}_3\text{S}$ ($T_c \sim 200\text{ K}$) and $\text{LaH}_{10}$ ($T_c \sim 280\text{ K}$)~\cite{duan2014pressure, peng2017hydrogen, liu2017potential}. However, these extraordinary transition temperatures are fundamentally bound to extreme megabar pressures (typically 150–180 GPa) required to stabilize the hydrogen covalent networks~\cite{drozdov2015conventional, einaga2016crystal, drozdov2019superconductivity, somayazulu2019evidence}. The massive technological barrier imposed by such pressures has shifted the overarching paradigm toward finding alternative chemical environments capable of sustaining high-$T_c$ BCS superconductivity at ambient or near-ambient conditions.

To bypass the requirement of extreme external compression, reducing the dimensionality of the host lattice has emerged as a highly promising route. Two-dimensional (2D) materials provide an ideal platform for exploring emergent quantum phenomena arising from reduced screening and enhanced many-body interactions. Atomically thin materials are known to host a rich variety of collective electronic states, including superconductivity, magnetism, topological phases, and charge-density waves (CDWs)~\cite{balandin2021charge}. Within these systems, the interplay between superconductivity and CDW order is of particular interest, as both phases emerge from electronic states near the Fermi level and are intensely sensitive to momentum-dependent EPC and selective phonon softening~\cite{wang2023interplay,ali2025interplay,johannes2008fermi,zhu2017misconceptions,jseeyang_cdw_moxh,seeyangnok2026triangular}. Understanding how EPC drives, stabilizes, or suppresses these competing collective states—a phenomenon observed in numerous layered compounds like NbSe$_2$, TaS$_2$, and recently predicted monolayers like Mo$_2$NF$_2$~\cite{ugeda2016characterization,xi2015strongly,lian2023interplay,hwang2024charge}—is critical for designing robust 2D superconductors.

In recent years, chemical functionalization has proven highly effective in engineering the electronic and vibrational properties of these 2D frameworks~\cite{bekaert2019hydrogen,seeyangnok2026enhanced,liu2024three,seeyangnok2026electron}. By modifying the local bonding environment, the incorporation of hydrogen into 2D lattices fundamentally reshapes the electronic and phononic density of states, heavily favoring EPC enhancement. This also has been demonstrated in in the 2D materials through substitutional hydrogenation. For example, these include Janus transition-metal chalcogenide hydrides such as MoSH, MoSeH, WSH, and related group-IV analogues~\cite{lu2017janus,liu2022two,sui2025two,seeyangnok2024superconductivity,qiao2024prediction,seeyangnok2024superconductivity,seeyangnok2025phase,seeyangnok2026tunable,seeyangnok2024superconductivitywseh}, calcium-intercalated bilayer graphene~\cite{seeyangnok2025hydrogenation}, transition-metal borides~\cite{seeyangnok2025high,seeyangnok2026stability}, and MXene monolayers~\cite{seeyangnok2025ab,seeyangnok2026theoretical}. In these systems, strong EPC gives rise to structural instabilities and phonon-mediated superconductivity with a wide range of transition temperatures~\cite{seeyangnok2026competition, ku2023ab,li2024machine}. 

Despite these successes, many substitutionally hydrogenated monolayers and Janus architectures exhibit complex magnetic ground states or broken inversion symmetries~\cite{seeyangnok2025competition} and Fermi instability of high electronic density of states~\cite{seeyangnok2025robust}. This introduces a rigorous interplay between magnetism and superconductivity that can ultimately suppress Cooper pairing. To maximize the benefits of a hydrogen-rich lattice without the complications of asymmetric magnetic phases, fully hydrogenated 2D transition-metal frameworks present an underexplored and highly promising frontier.

In this work, we investigate the structural stability, electronic properties, and phonon-mediated superconductivity of the hexagonal $T$-phase molybdenum dihydride ($T$-$\text{MoH}_2$) monolayer. While previous reports focused on the $H$-phase framework~\cite{yan2022enhanced}, our total-energy calculations reveal that the $T$-$\text{MoH}_2$ phase is energetically more favorable by $0.198\text{ eV}$, establishing it as the true ground-state configuration. Utilizing this symmetric, fully hydrogenated transition-metal lattice as a pristine model, we systematically evaluate the limits of hydrogen-derived EPC and investigate the presence of competing charge-density-wave instabilities in two dimensions. Our density functional perturbation theory calculations demonstrate that the dense hydrogen integration governs the emergence of robust conventional superconductivity. Driven by strong lattice dynamics, the $T$-$\text{MoH}_2$ monolayer exhibits an EPC parameter of $\lambda = 1.04$, yielding an isotropic resolved superconducting transition temperature ($T_c$) of $14.4\text{ K}$ at ambient pressure. These findings provide a crucial step forward in the realization of stable, ambient-pressure, 2D superconductors.

\section{Computational Methods}
Density functional theory (DFT) calculations were carried out using the \textsc{Quantum ESPRESSO} (QE) package~\cite{giannozzi2009quantum}. A plane-wave basis set was employed with a kinetic energy cutoff of 80~Ry for the wavefunctions and 320~Ry for the charge density. Brillouin zone integrations were performed using a \(24 \times 24 \times 1\) Monkhorst--Pack \textit{k}-point mesh~\cite{monkhorst1976special}. Electronic occupations were treated using the Methfessel--Paxton smearing scheme with a smearing width of 0.02~Ry~\cite{methfessel1989high}. The electron--ion interaction was described by optimized norm-conserving Vanderbilt pseudopotentials~\cite{hamann2013optimized,schlipf2015optimization}, while exchange--correlation effects were treated within the generalized gradient approximation (GGA) using the Perdew--Burke--Ernzerhof (PBE) functional~\cite{perdew1996generalized}.

Structural relaxations were performed using the BFGS algorithm~\cite{liu1989limited} until the residual forces on each atom were less than \(10^{-5}~\text{eV/\AA}\). Phonon spectra and dynamical properties were calculated within density functional perturbation theory (DFPT)~\cite{baroni2001phonons} on a \(12 \times 12 \times 1\) \textit{q}-point grid. The electron--phonon interaction leads to finite phonon linewidths $\gamma_{\boldsymbol{q}\nu}$, and the corresponding mode-resolved electron--phonon coupling strength is given by
\begin{equation}
\lambda_{\boldsymbol{q}\nu} = 
\frac{\gamma_{\boldsymbol{q}\nu}}{\pi N(\epsilon_F)\,\omega_{\boldsymbol{q}\nu}^2},
\end{equation}
where \( N(\epsilon_F) \) is the electronic density of states at the Fermi level and \(\omega_{\boldsymbol{q}\nu}\) denotes the phonon frequency. 

Superconducting properties were investigated by solving the anisotropic Migdal--Eliashberg equations~\cite{migdal1958interaction,eliashberg1960interactions,nambu1960quasi,pinsook2024analytic} using the EPW code~\cite{noffsinger2010epw,ponce2016epw}, which relies on Wannier--Fourier interpolation~\cite{giustino2007electron,giustino2017electron}. The superconducting gap function \(\Delta_{nk}(i\omega_j)\) and the renormalization function \(Z_{nk}(i\omega_j)\) were determined self-consistently given by 
The Migdal--Eliashberg equations are given by
\begin{eqnarray}
Z_{nk}(i\omega_j) &=& 1 + \frac{\pi T}{N(\varepsilon_F)\omega_j} 
\sum_{mk'j'} 
\frac{\omega_{j'}}{\sqrt{\omega_{j'}^2 + \Delta_{mk'}^2(i\omega_{j'})}} \nonumber \\
&&\times \lambda(nk, mk', \omega_j - \omega_{j'})\,
\delta(\epsilon_{mk'} - \varepsilon_F),
\end{eqnarray}
\begin{eqnarray}
Z_{nk}(i\omega_j)\Delta_{nk}(i\omega_j) &=& 
\frac{\pi T}{N(\varepsilon_F)} 
\sum_{mk'j'} 
\frac{\Delta_{mk'}(i\omega_{j'})}{\sqrt{\omega_{j'}^2 + \Delta_{mk'}^2(i\omega_{j'})}}  \delta(\epsilon_{mk'} - \varepsilon_F) \nonumber  \\ &&
\times \left[\lambda(nk, mk', \omega_j - \omega_{j'}) - \mu^*\right],
\end{eqnarray}
on the Matsubara frequency axis, where \(\omega_j = (2j+1)\pi T\). The Coulomb pseudopotential was fixed at \(\mu^* = 0.1\).

To ensure convergence, dense \textit{k}- and \textit{q}-point meshes of \(200 \times 200 \times 1\) and \(100 \times 100 \times 1\), respectively, were used. A Fermi surface broadening of 0.60~eV and a Matsubara frequency cutoff of 1.50~eV were applied. The delta functions were approximated using Gaussian smearing, with widths of 0.15~eV for electrons and 0.5~meV for phonons.

\section{Reselts and Discussion}

\subsection{Structural Stability}
    \begin{figure}[ht]
        \centering
        \includegraphics[width=8.5cm]{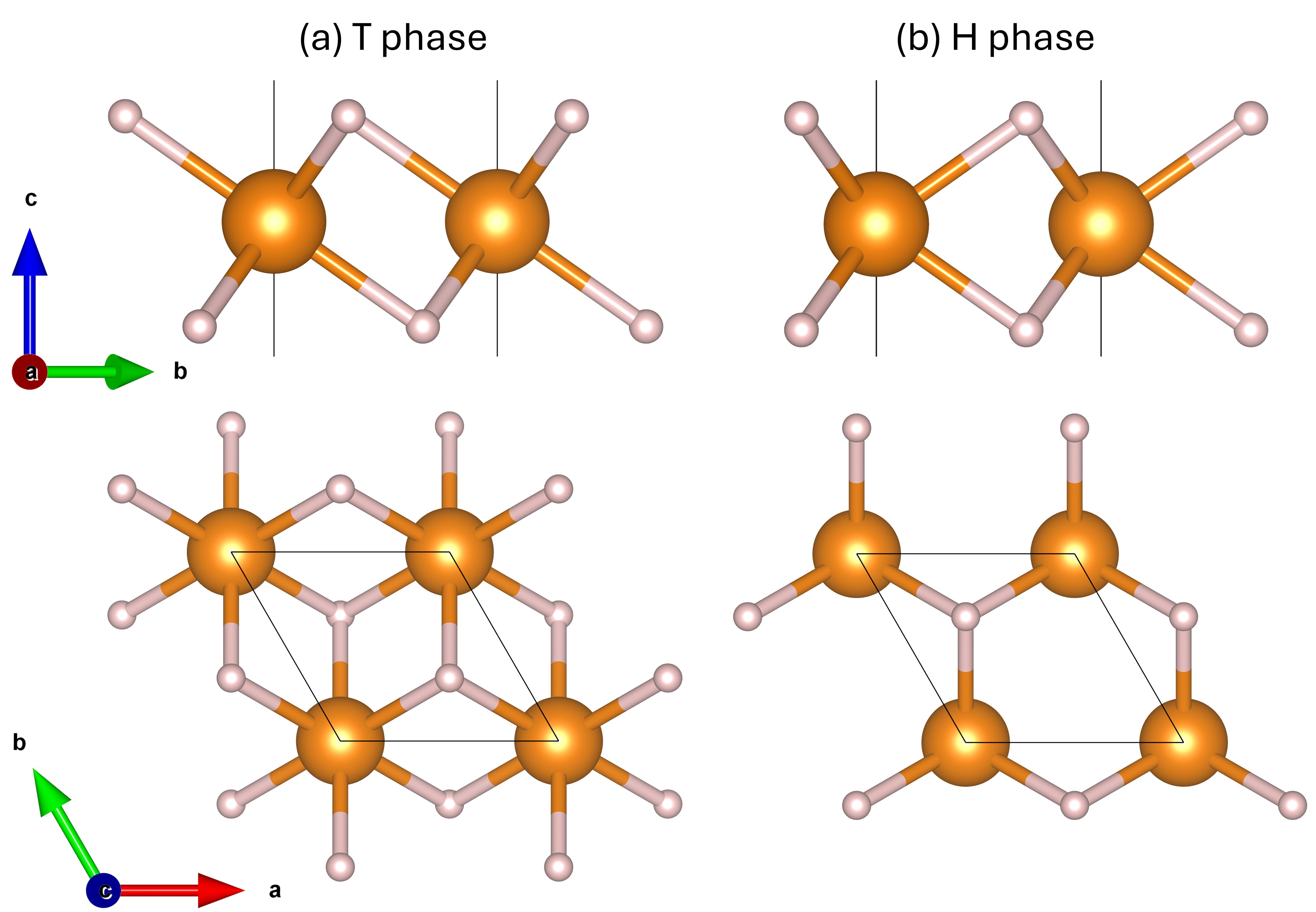}
        \caption{Crystal structures of monolayer molybdenum dihydride ($\text{MoH}_2$) in the (a) $T$-phase (octahedral coordination) and (b) $H$-phase (trigonal prismatic coordination). For each phase, the upper panel shows the side view along the $a$-axis (with the vertical thickness and $c$-axis indicated), and the lower panel shows the top view down the $c$-axis, where the unit cell is outlined by the black solid lines. Orange and pink spheres represent $\text{Mo}$ and $\text{H}$ atoms, respectively.}
        \label{fig:1}
    \end{figure}

    To understand the structural characteristics of the two competing polytypes of monolayer $\text{MoH}_2$, we systematically compare their relaxed lattice geometries. The $T$-phase crystallizes in a centrosymmetric structure with octahedral coordination [Fig.~\ref{fig:1}(a)], where the $\text{Mo}$ atoms are sandwiched between staggered layers of hydrogen atoms. In contrast, the $H$-phase exhibits a mirror-symmetric, trigonal prismatic coordination [Fig.~\ref{fig:1}(b)] featuring vertically aligned hydrogen layers. Our structural relaxation yields distinct differences in their respective lattice parameters. $T$-phase: exhibits a slightly larger in-plane lattice constant of $a = 2.76\text{ \AA}$ combined with a compressed vertical layer thickness (defined as the vertical distance between the upper and lower hydrogen planes) of $2.25\text{ \AA}$. $H$-phase: possesses a slightly smaller in-plane lattice constant of $a = 2.74\text{ \AA}$ but a significantly larger vertical thickness of $2.34\text{ \AA}$. Our calculations show that the $T$-phase is energetically more favorable than the $H$-phase by $0.198\text{ eV}$ per formula unit. This energy difference corresponds to a highly robust thermal energy barrier of approximately $2{,}296\text{ K}$. This significant energetic preference clearly establishes the $T$-phase as the true structural ground state of the $\text{MoH}_2$ monolayer. 
    
    The mechanical stability of the monolayer was established by determining the in-plane elastic constants through polynomial fitting of strain–energy curves. Small planar strains ($\epsilon_i, \epsilon_j$) were applied to the optimized unit cell, and the resulting total energies ($E$) were fitted to extract the stiffness coefficients, defined as:
    \begin{equation}
        C_{ij} = \frac{1}{S_0} \frac{\partial^2 E}{\partial \epsilon_i \partial \epsilon_j}
    \end{equation}
    with $S_0$ representing the unstrained cell area. Governed by hexagonal symmetry, the material exhibits two independent elastic parameters, $C_{11}$ and $C_{12}$, alongside the derived shear modulus $C_{66} = (C_{11} - C_{12})/2$. These constants dictate the linear elastic response via the generalized Hooke's law:
    \begin{equation}
        \sigma = \begin{bmatrix} C_{11} & C_{12} & 0 \\ C_{12} & C_{11} & 0 \\ 0 & 0 & C_{66} \end{bmatrix} \varepsilon .
    \end{equation}
    The mechanical stability of the hexagonal $T$-$\text{MoH}_2$ monolayer is confirmed by its calculated elastic constants: $C_{11} = 146.62\text{ N/m}$, $C_{12} = 80.45\text{ N/m}$, and $C_{66} = 33.08\text{ N/m}$. For a 2D hexagonal lattice ($C_{11} = C_{22}$), the Born stability criteria require $C_{11} > \vert{}C_{12}\vert{}$ and $C_{66} > 0$ to ensure a positive-definite strain energy density~\cite{mouhat2014necessary}. Because these moduli strictly satisfy these conditions, the monolayer is proven mechanically stable against uniform deformation and structural strain.

    \begin{figure}[ht]
        \centering
        \includegraphics[width=8.5cm]{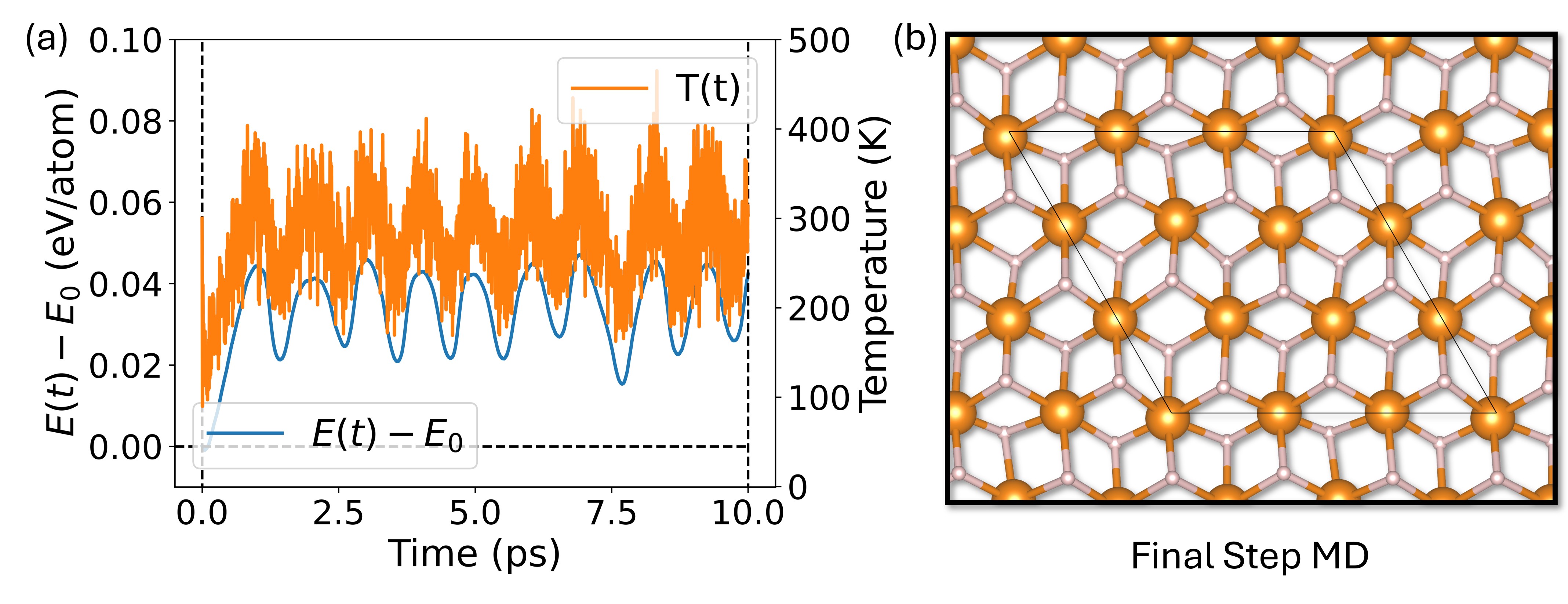}
        \caption{Thermal stability of the monolayer $T$-$\text{MoH}_2$ from ab initio molecular dynamics (AIMD) simulations. (a) Time evolution of the total energy fluctuation per atom, $E(t) - E_0$ (left axis, blue line), and temperature $T(t)$ (right axis, orange line) over a $10\text{ ps}$ simulation run at $300\text{ K}$. (b) Top-down snapshot of the $3 \times 3 \times 1$ supercell structure of 27 atoms at the final step ($10\text{ ps}$) of the AIMD simulation. Orange and pink spheres represent $\text{Mo}$ and $\text{H}$ atoms, respectively.}
        \label{fig:2}
    \end{figure}
    The thermal stability of the monolayer $T$-$\text{MoH}_2$ at $300\text{ K}$ was evaluated using AIMD simulations with a $3 \times 3 \times 1$ supercell of 27 atoms and a $1.0\text{ fs}$ time step over $10\text{ ps}$. The simulations used the NVT ensemble with a Nosé-Hoover thermostat~\cite{NHC} and a $1.0\text{ ps}$ coupling time constant. Consequently, the $\sim 1.0\text{ ps}$ oscillation period in the energy fluctuations [Fig.~\ref{fig:2}(a)] directly reflects this thermostat relaxation time rather than structural instability. As shown in Figure~\ref{fig:2}(a), both temperature and total energy fluctuate within well-defined bounds without systematic drift. Crucially, the final-step snapshot ($10\text{ ps}$) in Figure~\ref{fig:2}(b) demonstrates that the hexagonal lattice remains fully intact, exhibiting only typical thermal vibrations around equilibrium positions. No signs of lattice reconstruction, phase transitions, or hydrogen desorption are observed, confirming that the $T$-$\text{MoH}_2$ monolayer is thermodynamically stable at room temperature.
    
\subsection{Electronic Properties}
    \begin{figure}[ht]
        \centering
        \includegraphics[width=8.5cm]{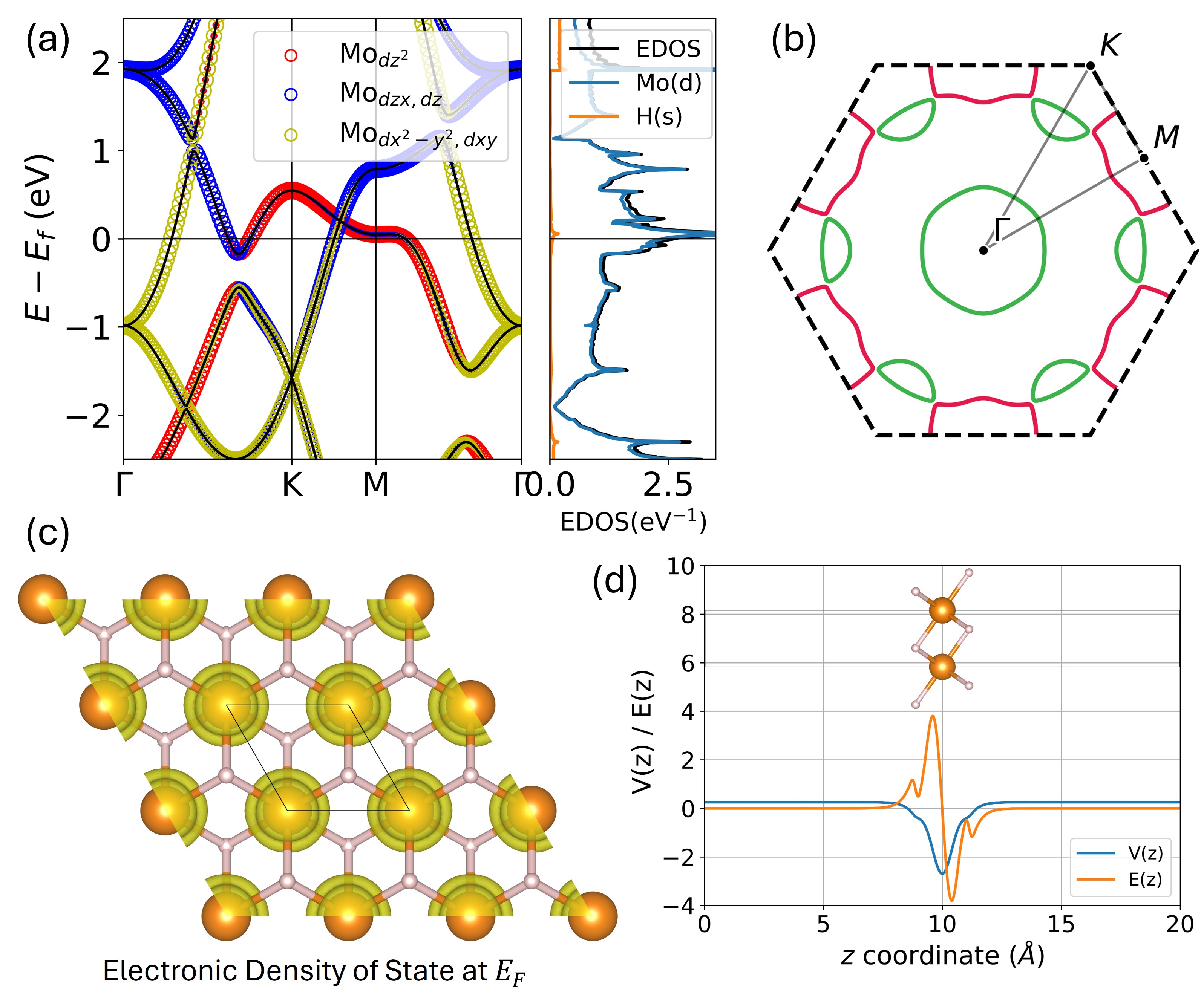}
        \caption{Electronic structure of the monolayer $T$-$\text{MoH}_2$. (a) Orbital-resolved electronic band structure along high-symmetry paths (left panel) and corresponding total and partial electronic density of states (EDOS, right panel). (b) Fermi surface topology across the first Brillouin zone, highlighting electronic pockets surrounding the $\Gamma$, $K$, and $M$ points. (c) Spatial distribution of the electronic charge density integrated within a narrow energy window around $E_F$. (d) Planar-averaged electrostatic potential $V(z)$ and local electric field $E(z)$ profiled along the vertical $z$-direction across the monolayer plane.}
        \label{fig:3}
    \end{figure}

    To elucidate the electronic ground state of the ground-state $T$-$\text{MoH}_2$ monolayer, we compute its electronic band structure, orbital-resolved density of states, and Fermi surface topology [Fig. ~\ref{fig:3}]. The band structure in Figure~\ref{fig:3}(a) demonstrates that the monolayer is inherently metallic, featuring multiple bands crossing the Fermi level ($E_F$). The states near $E_F$ are overwhelmingly dominated by the molybdenum $d$-orbitals, with the most significant contributions originating from the $\text{Mo-}d_{z^2}$ and $\text{Mo-}d_{x^2-y^2, d_{xy}}$ states. In contrast, the $\text{H-}s$ states exhibit negligible density near the Fermi level, primarily contributing to deeper valence bands. This indicates that the metallicity is primarily driven by the transition-metal framework.

    The multi-band metallic nature of $T$-$\text{MoH}_2$ yields a complex, well-defined Fermi surface [Fig. ~\ref{fig:3}(b)]. The topology consists of distinct concentric pockets centered at the $\Gamma$ point, alongside additional pockets located near the $K$ and $M$ high-symmetry points. The spatial distribution of the electronic states at $E_F$ [Fig.~\ref{fig:3}(c)] further confirms that the conducting electrons are heavily localized around the $\text{Mo}$ atomic sites, forming a continuous in-plane conducting network. Finally, the planar-averaged electrostatic potential and local electric field plotted along the $z$-axis [Fig.~\ref{fig:3}(d)] confirm structural inversion symmetry, showing perfectly symmetric potential barriers centered on the monolayer plane. The substantial electronic density of states as shown in Figure~\ref{fig:3} (a) at the Fermi level ($N(E_F)$), dominated by localized $\text{Mo}$-$d$ states, provides a highly favorable electronic environment for mediating strong conventional electron-phonon coupling.

\subsection{Phonon-Mediated Superconductivity}
    \begin{figure}[ht]
        \centering
        \includegraphics[width=8.5cm]{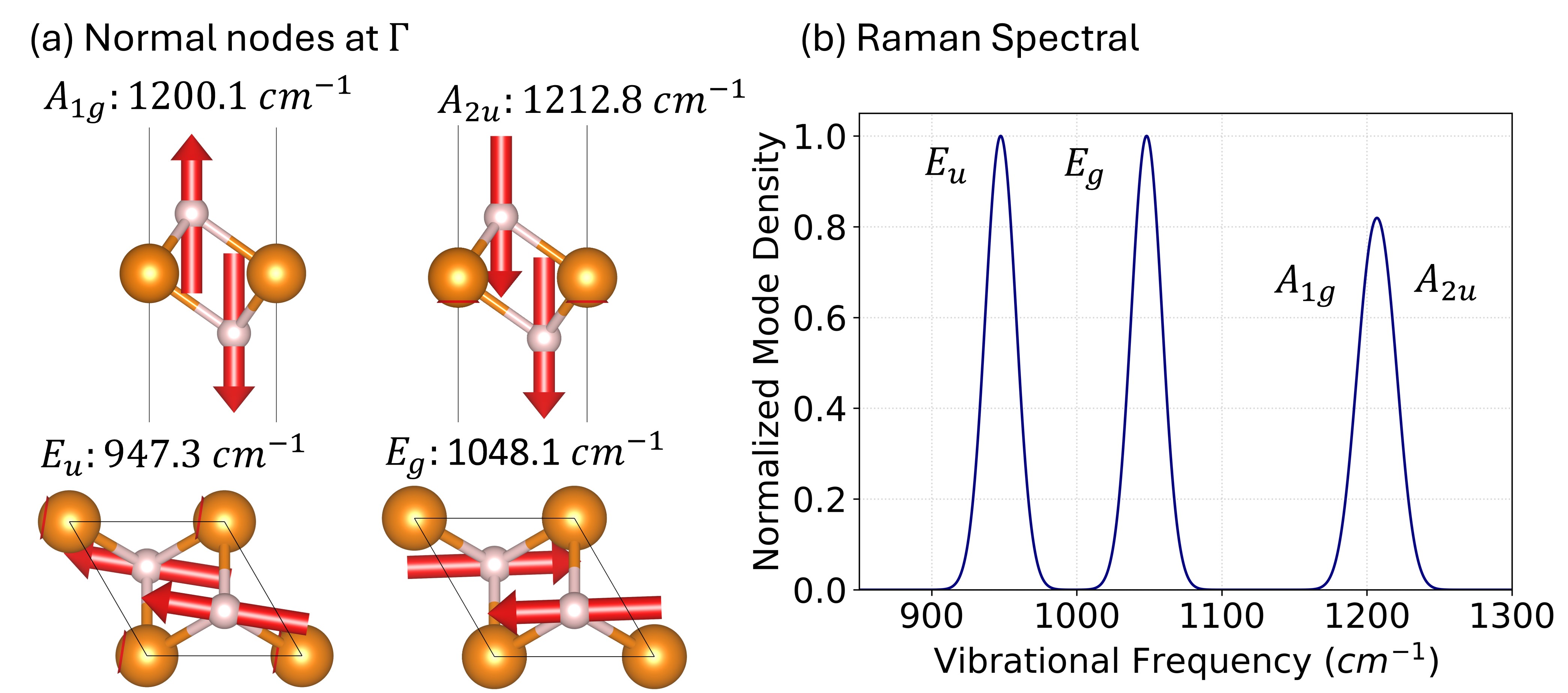}
        \caption{Vibrational modes and simulated Raman spectrum of monolayer $T$-$\text{MoH}_2$ at the Brillouin zone center ($\Gamma$). (a) Schematic representation of the atomic displacement vectors for the four distinct optical phonon modes: $E_u$ ($947.3\text{ cm}^{-1}$), $E_g$ ($1048.1\text{ cm}^{-1}$), $A_{1g}$ ($1200.1\text{ cm}^{-1}$), and $A_{2u}$ ($1212.8\text{ cm}^{-1}$). Red arrows denote the direction and relative amplitude of atomic motions. (b) Smeared phonon mode profile representing the simulated high-frequency vibrational spectrum, with peaks labeled according to their corresponding irreducible representations.}
        \label{fig:4}
    \end{figure}

    To characterize the lattice dynamics of the ground-state $T$-$\text{MoH}_2$ monolayer, we analyze its optical phonon modes at the zone center ($\Gamma$) [Fig.~\ref{fig:4}]. Group theory analysis dictates that the optical modes span the irreducible representations of the $D_{3d}$ point group. The calculated high-frequency spectrum displays four distinct optical features above $900\text{ cm}^{-1}$, which originate entirely from the lightweight hydrogen networks.

    As illustrated in Figure~\ref{fig:4}(a), the lower-frequency optical modes exhibit in-plane displacements. The doubly degenerate $E_u$ mode at $947.3\text{ cm}^{-1}$ corresponds to the in-phase in-plane motion of the upper and lower hydrogen layers, whereas the doubly degenerate $E_g$ mode at $1048.1\text{ cm}^{-1}$ represents their anti-phase stretching vibration against the fixed molybdenum sublattice. At higher frequencies, the modes are dominated by out-of-plane movements. The $A_{1g}$ mode at $1200.1\text{ cm}^{-1}$ consists of the symmetric, out-of-phase vertical stretching of the hydrogen atoms away from the central metal plane. Conversely, the $A_{2u}$ mode at $1212.8\text{ cm}^{-1}$ involves in-phase vertical translation of both hydrogen layers relative to the Mo plane. The simulated mode density profile [Fig.~\ref{fig:4}(b)] clearly resolves these features, providing distinct spectral fingerprints that can be directly verified via future experimental Raman or infrared spectroscopy measurements to confirm the structural realization of the $T$-$\text{MoH}_2$ monolayer.
    
    \begin{figure}[ht]
        \centering
        \includegraphics[width=8.5cm]{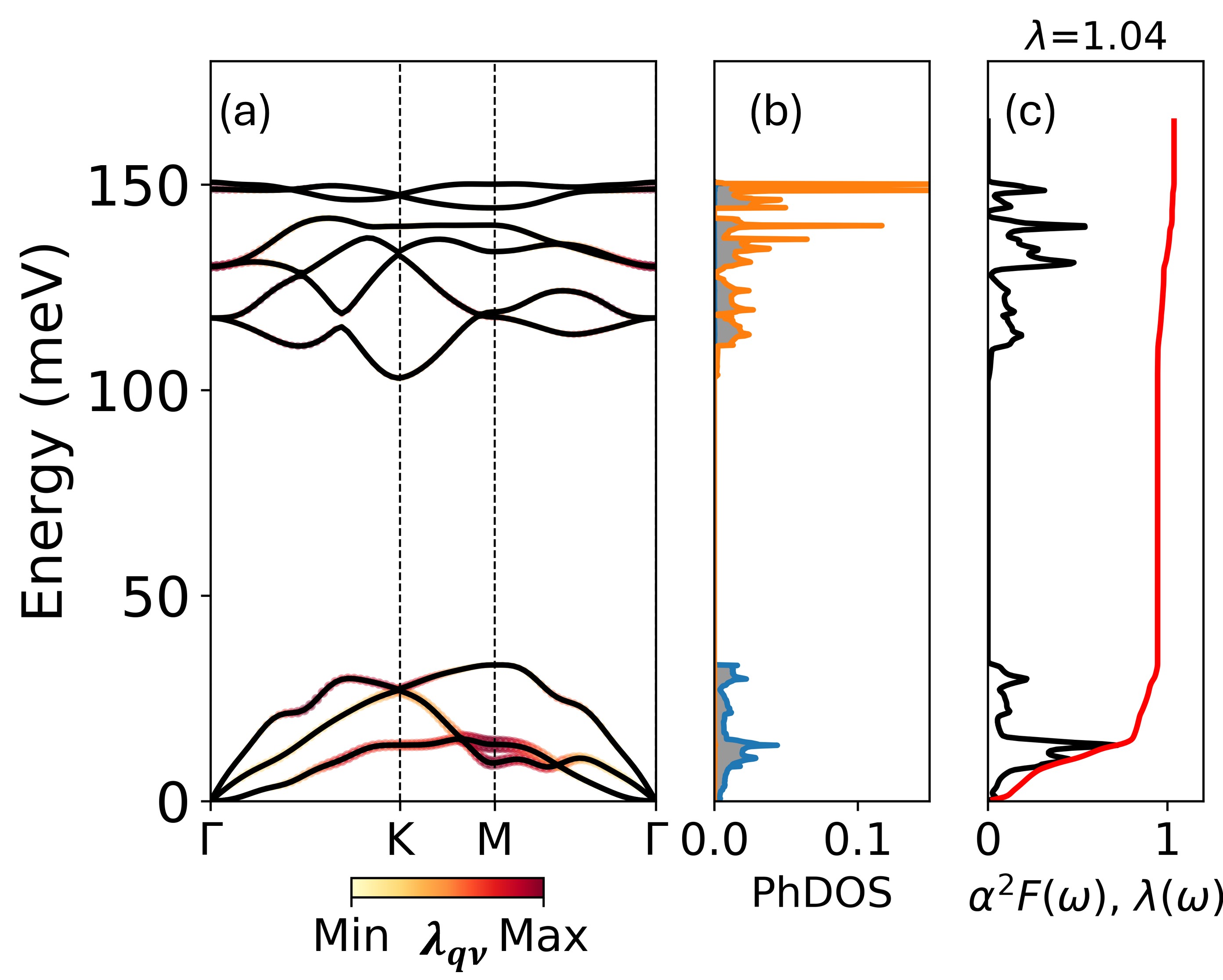}
        \caption{Phonon dispersion relations and electron-phonon coupling (EPC) properties of monolayer $T$-$\text{MoH}_2$. (a) Phonon band structure along high-symmetry paths, where the color mapping indicates the strength of the mode-resolved EPC linewidth ($\gamma_{\mathbf{q}\nu}$). (b) Phonon density of states (PhDOS) projected onto the constituent $\text{Mo}$ (blue) and $\text{H}$ (orange) atoms. (c) The Eliashberg electron-phonon spectral function $\alpha^2F(\omega)$ (black line) and the cumulative frequency-dependent EPC parameter $\lambda(\omega)$ (red line), yielding a total integrated value of $\lambda = 1.04$.}
        \label{fig:5}
    \end{figure}

    To evaluate the conventional phonon-mediated superconductivity in the $T$-$\text{MoH}_2$ monolayer, we conduct density functional perturbation theory (DFPT) calculations to analyze its lattice dynamics and electron-phonon coupling (EPC) properties [Fig.~\ref{fig:5}]. The phonon dispersion relations displayed in Figure~\ref{fig:5}(a) exhibit no imaginary frequencies across the entire Brillouin zone, confirming the robust dynamical stability of the $T$-phase crystal structure. The spectrum splits into two well-separated regions: a low-energy acoustic and interstitial domain below $35\text{ meV}$ dominated by heavy $\text{Mo}$ vibrations, and a high-energy optical band between $100\text{ meV}$ and $150\text{ meV}$ originating exclusively from the lightweight $\text{H}$ atom networks [Fig.~\ref{fig:5}(b)].

    The mode-resolved EPC strength, mapped onto the phonon branches in Figure~\ref{fig:5}(a), reveals that the low-frequency acoustic modes (particularly along the $K$-$M$-$\Gamma$ path) exhibit substantial coupling to the electronic states near the Fermi level. This contribution is clearly visible in the Eliashberg spectral function $\alpha^2F(\omega)$ [Fig. 5(c)], which displays prominent peaks in the low-energy regime. Consequently, the cumulative integration of $\lambda(\omega)$ rises steadily across the acoustic range to approximately $0.95$ , which accounts for 91\% of the total integrated EPC parameter of $\lambda = 1.04$.

    \begin{figure}[ht]
        \centering
        \includegraphics[width=8.5cm]{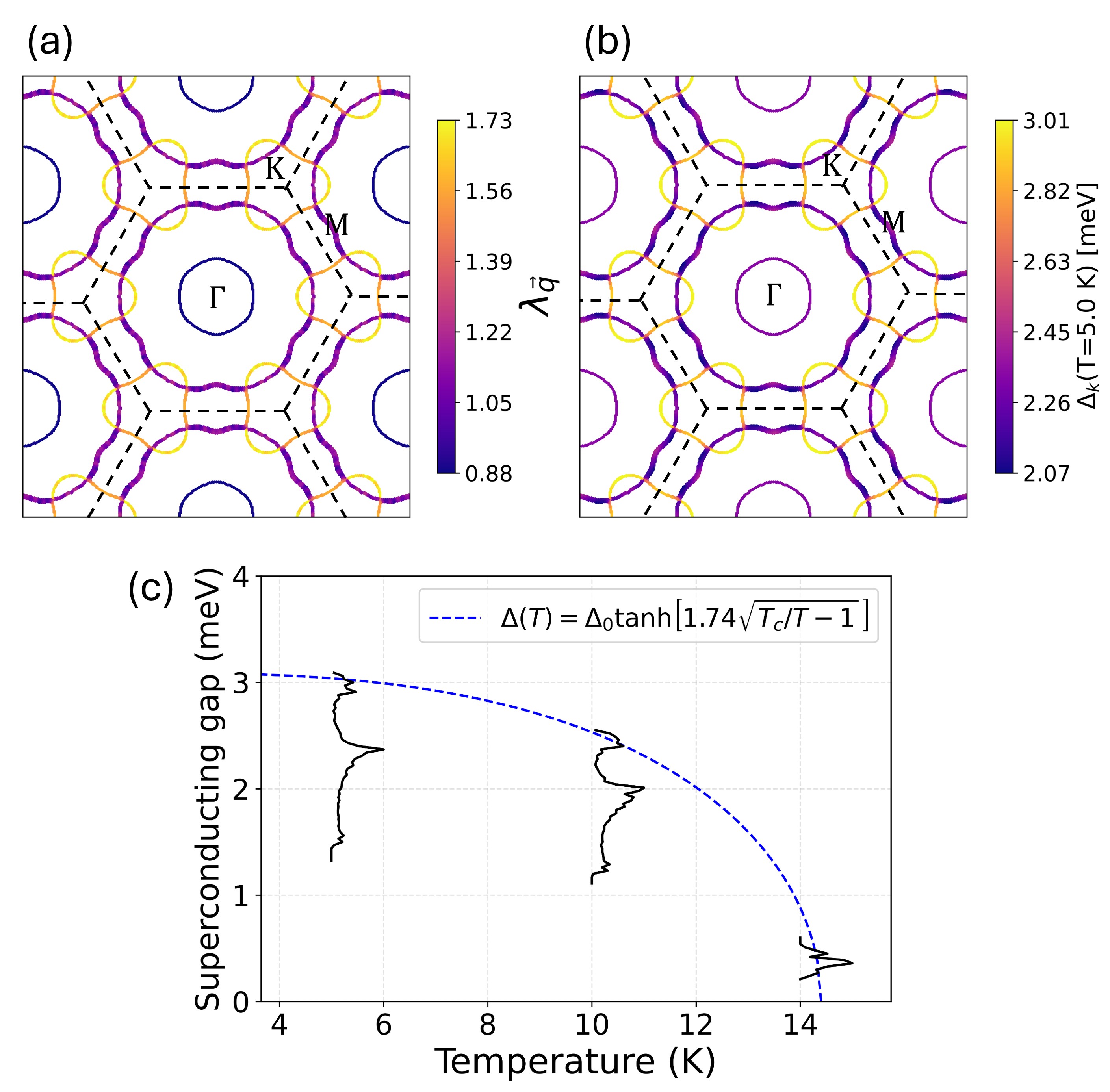}
        \caption{Momentum-dependent superconducting properties and gap evolution of monolayer $T$-$\text{MoH}_2$ solved via the anisotropic Migdal-Eliashberg formalism. (a) Momentum-resolved electron-phonon coupling parameter $\lambda_{\mathbf{k}}$ mapped onto the Fermi surface sheets across the first Brillouin zone. (b) Momentum-dependent superconducting gap function $\Delta_{\mathbf{k}}$ evaluated at $T = 5.0\text{ K}$. (c) Temperature-dependent evolution of the superconducting gap $\Delta(T)$ showing the distribution of gap values (black curves) closing at a critical temperature ($T_c$) of $14.4\text{ K}$, compared against the conventional BCS isotropic fit (blue dashed line).}
        \label{fig:6}
    \end{figure}

    To capture the momentum-dependent features of the superconducting state in the $T\text{-MoH}_2$ monolayer, we solve the anisotropic Migdal-Eliashberg equations. Figure~\ref{fig:6}(a) illustrates the momentum-resolved electron-phonon coupling parameter $\lambda_{\mathbf{k}}$ mapped directly onto the Fermi surface. The coupling exhibits a notable variation across the Fermi sheets; specifically, the sheets with dominant $d_{zx,zy}$ character around the $K$ point exhibit a strong EPC that reaches an upper maximum of $1.73$, whereas the other Fermi sheets display a weaker coupling that continuously drops to a lower limit of $0.88$

    This momentum-dependent coupling directly translates into an isotropic superconducting gap distribution $\Delta_{\mathbf{k}}$ at low temperatures ($T = 5.0\text{ K}$), as shown in Figure~\ref{fig:6}(b). The superconducting gap varies continuously between $2.07\text{ meV}$ and $3.01\text{ meV}$, tracking the profile of the underlying EPC strength. The temperature-dependent evolution of this gap distribution is plotted in Figure~\ref{fig:6}(c). As the temperature increases, the spreadsheet of gap values steadily contracts and eventually closes at a critical transition temperature ($T_c$) of $14.4\text{ K}$. The calculated gap closing follows the general trend of the standard BCS-like thermodynamic curve, confirming that the conventional, phonon-mediated pairing mechanism in $T$-$\text{MoH}_2$ maintains a robust, moderately anisotropic multi-band superconducting state.
\section*{Conclusions}
    In summary, we have systematically investigated the structural stability, electronic properties, and conventional phonon-mediated superconductivity of the hexagonal $T$-phase molybdenum dihydride ($T\text{-MoH}_2$) monolayer using first-principles calculations. Total-energy evaluations demonstrate that the octahedral $T$-phase is the true thermodynamic ground-state configuration, being energetically more favorable than the previously reported trigonal prismatic $H$-phase by $0.198\text{ eV}$ per formula unit ($2{,}296\text{ K}$). The structural integrity and experimental feasibility of this 2D structure are firmly validated by its elastic constants, which strictly satisfy the Born mechanical stability criteria , and by ab initio molecular dynamics (AIMD) simulations confirming robust thermal stability at room temperature ($300\text{ K}$) without any lattice reconstruction or hydrogen desorption.

    Electronic structure calculations reveal that the $T\text{-MoH}_2$ monolayer is inherently metallic. The states near the Fermi level are predominantly governed by localized $\text{Mo}$-$d$ orbitals, which form a multi-band Fermi surface. Driven by the integration of the lightweight hydrogen network, density functional perturbation theory (DFPT) calculations demonstrate strong electron-phonon coupling (EPC) within the acoustic modes, yielding a substantial total integrated EPC parameter of $\lambda = 1.04$. By solving the anisotropic Migdal-Eliashberg equations, we determine that the superconducting gap exhibits an isotropic distribution ($\Delta = 2.07\text{--}3.01\text{ meV}$ at $5.0\text{ K}$) and successfully closes at a critical transition temperature ($T_c$) of $14.4\text{ K}$ under ambient pressure. These findings highlight the fully hydrogenated $T\text{-MoH}_2$ monolayer as a promising, stable platform for exploring conventional 2D superconductivity under ambient conditions, bypassing the extreme pressure requirements of bulk hydrides.

\section*{Acknowledgments}
	This research project is supported by the Second Century Fund (C2F), Chulalongkorn University (Grant No. C2F PD-2320260067). High-performance computing facility in this Research is funded by Thailand Science research and Innovation Fund Chulalongkorn University (ST690022300001).

\bibliography{references}

\end{document}